\documentclass[12pt,preprint]{aastex}

\shorttitle{Young SSCs in M82: The Catalogue}
\shortauthors{Melo et al.}

\begin{document}

\title{Young Super Star Clusters in the Starburst of M82: The 
Catalogue\footnote{Based on 
observations made with the NASA/ESA {\itshape Hubble Space Telescope}, obtained 
from the data archive at the Space Telescope Science Institute. STScI is 
operated by the Association of Universities for Research in Astronomy, Inc., 
under NASA contract NAS 5-26555.}}

\author{V. P. Melo and C. Mu\~noz-Tu\~n\'on}
\affil{Instituto de Astrof\'{\i}sica de Canarias,
              V\'{\i}a L\'actea s/n, 38200 La Laguna,
              Spain}
\email{vmelo@iac.es,cmt@iac.es}

\author{J. Ma\'{\i}z-Apell\'aniz\altaffilmark{2}}
\affil{Space Telescope Science Institute, 3700 San Martin Drive,
	     Baltimore, MD 21218, USA}
\altaffiltext{2}
{Affiliated with the Space Telescope Division of the European Space 
Agency, ESTEC, Noordwijk, The Netherlands.}
\email{jmaiz@stsci.edu}

\and

\author{G. Tenorio-Tagle}
\affil{Instituto Nacional de Astrof\'{\i}sica \'Optica y Electr\'onica,
	     AP 51, 72000 Puebla, M\'exico}
\email{gtt@inaoep.mx}

\begin{abstract}
Recent results from {\itshape Hubble Space Telescope} ({\itshape HST\/})
have resolved starbursts as collections of compact
young stellar clusters. Here we present a photometric catalogue of the young 
stellar clusters in the nuclear starburst of M82, observed with the {\it HST} 
WFPC2 in H$\alpha$ (F656N) and in four optical broad-band filters. We identify 
197 young super stellar clusters. The compactness and high density of the 
sources led us to develop specific techniques to measure their sizes. 
Strong extinction lanes 
divide the starburst into five different zones and we provide a catalogue of 
young
super star clusters for each of these. In the catalogue we include relative 
coordinates, radii, fluxes, luminosities, masses, equivalent widths,
extinctions, and other parameters. Extinction values have been derived
from the broad-band images. The radii
range between 3 and 9~pc, with a mean value of 5.7~$\pm$~1.4~pc, 
and a stellar mass between 10$^4$ and 10$^6$ $M_\odot$. 
The inferred masses and mean separation, comparable to the 
size of super star clusters, together with their high volume density,
provides strong evidence for the key ingredients postulated by 
\citet*{Tenorio03}
as required for the development of a supergalactic wind.
\end{abstract}


\section{Introduction} \label{section1}

The {\itshape Hubble Space Telescope} has revealed star formation in starburst 
as
collections of centrally condensed super star clusters (SSCs) also referred to
as young globular clusters. SSCs are young star 
clusters with very high luminosity and compactness \citep{Ho97}. 
 Their typical sizes \citep[1--6~pc,][]{Meurer95,Maiz01} and masses 
($\sim$3~$\times$~10$^4$--10$^6$~$M_\odot$) are similar to those of globular 
clusters and they present a typical bolometric luminosity in the 
range 10$^{40}$--10$^{42.5}$ \citep{Strickland99}.
SSCs seem to be protoglobular clusters; they have a large mass and 
a Salpeter-type initial mass function (IMF), although other 
possibilities
have also been considered \citep[see][ and references therein]{Kaaret04}.
SSCs have been found in interacting galaxies 
\citep[the Antennae; ][]{Whitmore99}, in starburst galaxies 
\citep[NGC~253; ][]{Watson96}, and also in star formation regions in normal 
spiral galaxies \citep{Larsen99a}. For an exhaustive review see
the proceedings of the recent meeting {\itshape The Formation and Evolution of 
Massive Young Star Clusters}, held in November 2003 (to be published by ASP).

M82 provides the best example of a supergalactic wind (SGW) in the local
Universe. It has a biconical extended filamentary structure, as evidenced by 
Subaru 
\citep{Ohyama02}, embedded in a pool of soft X-ray emission as detected by
{\itshape Chandra} \citep{Griffiths00} and {\itshape 
XMM--Newton} \citep{Stevens03} and 
extending several kpc away from the nuclear region.
Moreover, its proximity made M82 a target for a large number of observational 
programs with {\itshape HST}. The observations  now available led us to go into 
the analysis of the SSC candidates with the aim of measuring the physical 
properties of a collection of SSCs able to develop and sustain a
supergalactic wind.

M82 is nearby and luminous; its H$\alpha$ luminosity is 
1.07~$\times$~10$^{41}$~erg s$^{-1}$ \citep{Lehnert96} after correction for 
galactic extinction. In this paper a distance of 3.6~Mpc, or m - M = 27.8~mag, 
is assumed, based on the Cepheid distance to M81 obtained by \cite{Freedman94}. 
The corresponding linear scale is 1$''$~=~17.5~pc.

The total extinction in the central regions of M82 is very high and patchy, 
and is mainly caused by a number of dark lanes crossing it 
(see Figure~\ref{fig:fig1}), which has led to a wide range of 
extinction values in the
literature. For example, \cite{Oconnell78} used an internal extinction of 
$A_V=2.5$~mag but more recent studies 
\citep*[see, e.g.,][]{Watson84,Oconnell95,Satyapal95,Alonso03} give extinction 
values in the range 2.5~mag $< A_V <$ 25~mag. 

Although the {\itshape HST} data archive covers the whole central starburst of 
M82,
only the region M82-B \citep*{deGrijs01,Parmentier03,deGrijs03} 
and the cluster M82-F \citep{Smith01} have been studied in detail.
M82-B is also known as the ``fossil starburst'' of M82 where 
\cite{deGrijs01} identified  super star cluster candidates
after analyzing {\itshape HST} images (WFPC2). \cite{deGrijs03} 
also found in this region a peak formation epoch at $\sim$1100 Myr for a 
sub-sample of clusters with well-determined ages.

In this paper we present a catalogue of young SSCs in the active starburst of 
M82,
the first detailed catalogue of the region, with relative positions and
parameters, such as luminosities, compactness and number of super star 
clusters, relevant to study the evolution
of a starburst. In Section~\ref{section2} we describe the data processing and 
analysis. Section~\ref{section3} describes the procedure 
developed to detect and measure the properties of SSC candidates.
Section~\ref{section4} presents the physical properties of the young clusters, 
and
a discussion of the results is given in Section~\ref{section5}.

\section{Data Calibration and Analysis of HST/WFPC2 Images}\label{section2}

The data were retrieved from the {\itshape Hubble Space Telescope} public data
archives and correspond to three different observational programs (see 
Table~\ref{tab1}). In this paper we analyze the H$\alpha$  
(F656N) images. Nitrogen~{\scshape ii} images 
([N~{\scshape ii}]~$\lambda$6583.6~\AA---F658N) are used to eliminate the 
contribution of this line in 
the H$\alpha$ images. Some broad-band images (F547M---$Str\ddot{o}mgren$ $\it y$
and F814W---I filter) are used to eliminate the contribution from the
background continuum. On the other hand, all broad-band images are used to
calculate the extinction and masses of SSCs (section~\ref{section3.4}).
Images are retrieved already processed by the 
standard WFPC2 pipeline. The calibration factor is obtained with the
STSDAS package {\scshape synphot}. Each field is observed in four exposures, 
which
allows us to combine them to eliminate cosmic rays (the {\scshape crrej} task
in the {\scshape stsdas} package). The calibration precision of the photometry
is better than 5~$\%$ \citep{Biretta02}.


The WFPC2 comprises four  800 $\times$ 800 pixel cameras. Three of these 
(the Wide Field chips or WF) have a  scale of
0.1$''$/pixel and the fourth (the Planetary Camera or PC) has
 a scale of 0.046$''$/pixel. For more 
information see \cite{Biretta02}.
We have used the PC images whose fields entirely cover the starburst area of 
M82 in all filters but F814W, F439W and F555W; these are covered by the WF4
camera. In order to match the PC pixel size, the F814W image has been 
oversampled. The geometric distortion was corrected using the 
{\scshape drizzle} task. 

The next step in the data processing is to subtract the continuum  
contribution from the H$\alpha$ emission. Ideally, one would use one or two 
narrow-band filters adjacent in wavelength to H$\alpha$ to estimate the 
continuum. In practice, such filters are usually not available and one has to 
use either (a) a wide-band filter close in wavelength but contaminated by 
emission lines or (b) two encompassing filters at a certain distance in 
wavelength, from where one can interpolate the continuum at H$\alpha$. Here, 
we follow the second approach since there is no $R$ band image available in 
the archive. For the continuum at wavelengths longer than H$\alpha$ there is 
only one filter available, F814W. In the blueward side, we can choose between
F555W and F547M. F555W is wider and provides a better S/N but it has one 
severe problem: it is contaminated by [\ion{O}{3}] $\lambda$4959+5007
and, to a lesser degree, by H$\alpha$ itself and H$\beta$. For that reason we
selected F547M as our blueward filter. We interpolated in wavelength the 
spectrum
between F547M and F814W choosing a synthetic spectra of {\scshape 
synphot} \citep{White98} 
in IRAF \citep{Tody86} for an atmosphere of a star with $T=30\,000$~K 
\citep{Kurucz79} 
whose ($F547M-F814W$) color matches the one  measured. 
Finally, we selected the value of the spectrum at the  wavelength of H$\alpha$ 
to estimate the continuum contribution to  F656N.

We also estimated the charge transfer efficiency corrections ($\pm$1
percent) and considered  these  not to be necessary. Finally,
the images were rotated to set them at the standard orientation (north up and
east left).

After calibration and continuum subtraction of the WFPC2 data, we compared our 
results with previous work. \cite{Oconnell78} measured the H$\alpha$ 
luminosity in all of M82-A and corrected for reddening (A$_V$ = 2.5), they 
obtained a luminosity of 5.06~$\times$~10$^{40}$erg~s$^{-1}$ (the difference in 
distance has been allowed for). Correcting for the same reddening and inside a
contour of 3~$\times$~10$^{-17}$ erg~s$^{-1}$~cm${-2}$, we measured 
5.02~$\times$~10$^{40}$ erg~s$^{-1}$, confirming the WFPC2 data calibration.


\section{Detection and Measurement of Candidate SSCs} \label{section3}

\begin{figure}
\epsscale{0.6}
\plotone{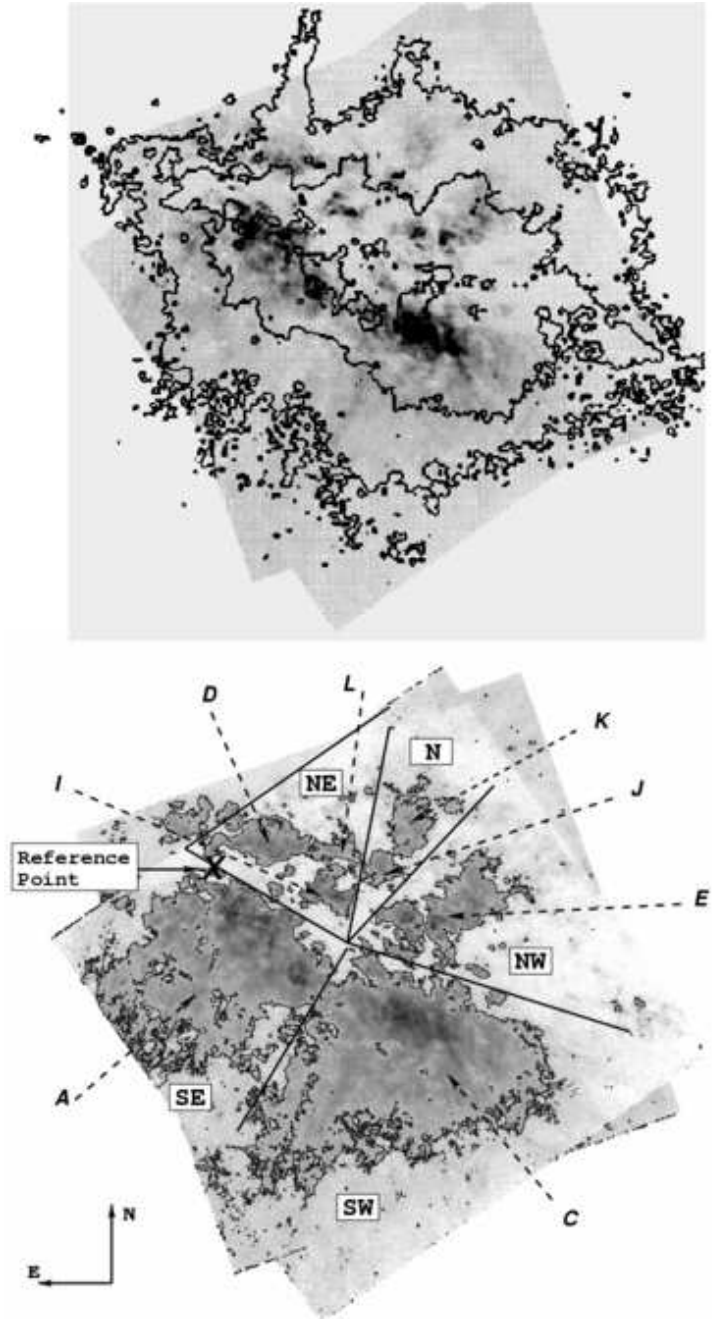}
\caption{The central region of M82. Upper image: WFPC2 H$\alpha$ image with
the NICMOS Fe~{\scshape ii} isocontours \citep{Alonso03} emission superimposed. 
The lower image shows the zones seen in the WFPC2 
H$\alpha$ image used in this paper (the names are indicated in white boxes). The 
nomenclature propose by \cite{Oconnell78} is indicated outside the image with 
dashes arrows. \label{fig:fig1}}
\end{figure}

The starburst of M82 is crossed by dark lanes which divide
it into five different zones (see Figure \ref{fig:fig1}, 
lower image), here used as reference to locate the SSCs in the nucleus of M82.
The figure (top panel) also shows the {\itshape HST}/WFPC2 image 
overlaid with isocontours of the {\itshape HST}/NICMOS [Fe~{\scshape ii}]1.644 
$\mu$m image \citep{Alonso03}. As the [Fe~{\scshape ii}] emission is not as 
affected by extinction, it allows a deeper view into the otherwise dark lanes 
in the visible, and this reveals the starburst as a continuous extended region. 
Note also how the [Fe~{\scshape ii}] emission follows the structure of the 
filaments seen in the visible, particularly in the northern part. 

The correspondence between the five zones (named in Figure~\ref{fig:fig1}, 
lower panel) and  the \cite{Oconnell78} sequence of bright regions in M82, 
extended as to 
also include the smaller regions (I - L) 
is given below:

\begin{description}

\item[N]:  zone to the north (J+K)
\item[NE]: zone to the northeast (I+D+L)
\item[NW]: zone to the northwest (E)
\item[SE]: zone to the southeast (A)
\item[SW]: zone to the southwest (C)

\end{description}

Note that region B, or the fossil starburst of M82 \citep{deGrijs01}, 
is to the east of the
starburst and outside the PC field. 

The zones evidently host a plethora of compact knots, with a variety of sizes.
However, not all bright H$\alpha$ knots can be associated with SSCs. They may 
result from H~{\scshape ii} regions ionized by an embedded young super star 
cluster, or could also be dense clouds illuminated from  the outside by nearby 
clusters. The inhomogeneous extinction is also important, 
since it  limits the number 
of observed SSCs in a manner difficult to estimate.

To discriminate among the various possibilities, three independent analyses 
of the images were carried out. First we consider the
error values in equivalent width, second we look for holes in the 
extinction map, and third we thoroughly analyze the continuum image. We 
identified bright knots both in H$\alpha$ and continuum images and selected
as young super star cluster candidates, those which are detected in both images.
Knots that only emit in H$\alpha$ are most probably illuminated clouds, whereas 
those that only show up in the continuum were declared old clusters; neither 
of these types are included in the catalogue.

\subsection{Search for bright knots} 

Bright knots were selected using {\scshape daofind} \citep{Stetson87}, 
as implemented in IRAF and using a feature detection threshold of 15-sigma.
{\scshape daofind} was optimized for the identification of pointlike sources, 
(roundish 
structures), what led us to consider it as suitable for finding compact SSCs.
However, few of them ($\leq$ 3\%), not being circular enough, are missed by the 
software. The 
``amorphous'' SSCs are either more than 
one SSC superposed in projection or clusters suffering channeled outflows of
ionized material. All the SSCs with emission outside
the roundish structures were followed and marked with an asterisk in the 
catalogue to indicate the cases where a lower limit to the measured flux is 
given.

\begin{figure}
\epsscale{0.5}
\plotone{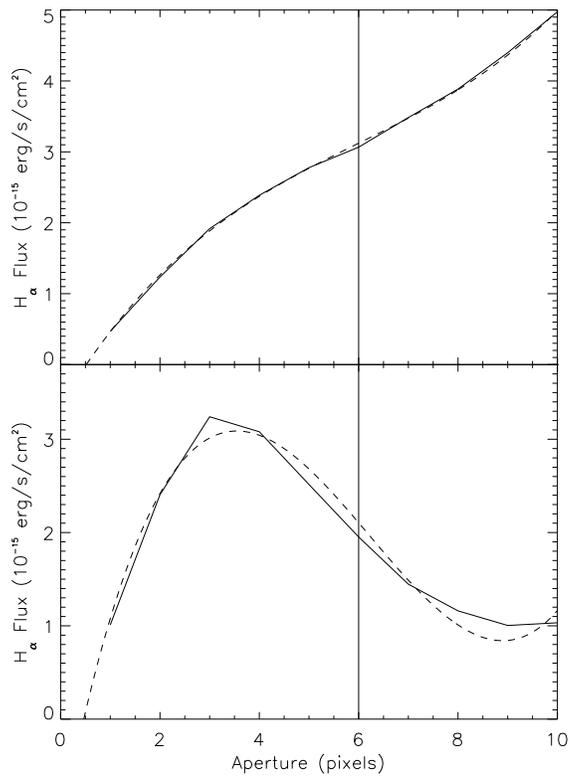}
\caption{Two examples of the method used to find the boundaries of knots. 
The solid line shows the differential flux profile for concentric annulus at 
different radius (1--10 pixels) and the dashed line is the third-order
polynomial. 
Inflection points are marked with a vertical line.
\label{fig:fig2}}
\end{figure}

Specific software was developed to identify and measure SSC 
candidates
in nearby galaxies using ground-based observations \citep{Larsen99b}. However, 
the
case we are analyzing is different. The problem is not PSF and seeing
deconvolution but crowding, and to define an algorithm able to identify and
measure a high density of sources such as those present 
in M82 and other starbursts.
To ascertain the {\it radii of SSCs} is not an 
easy task because of crowding in some areas;
there are many SSCs in a very 
small region so their fluxes are mixed. The procedure we have designed is as 
follows. From the already identified maxima, we made concentric 
apertures of increasing radius ($\Delta r$ = 1 pixel), from 1 to 10 
pixels, using the {\scshape phot} task in {\scshape apphot} (IRAF) 
(equivalent to 0.81--8.05~pc) and compute the flux profile for concentric
annuli at different radii both for H$\alpha$ and for the continuum.
We fit the resulting flux distribution of each knot with
a third-order polynomial and calculate the inflection points. In this way we 
establish the size and the overlap limit with the nearest neighbors. 
For the case of clusters with very close companions,
the estimated radius is a lower limit that clearly 
depends on the crowding in this region. In Figure~\ref{fig:fig2} two 
different cases are shown. The upper graph is a difficult 
case because the profile
is very monotonic due to crowding. The lower curve represents an isolated
SSC where it is relatively easy to find the limit of the region.
The method works properly in most cases.
Tables~\ref{tab3} and \ref{tab4} (see also section~\ref{section4})
list the emission peak centers and the estimated angular sizes, from
both H$\alpha$ and the 
continuum images, respectively. 
The complete tables are on-line in the electronic version of this paper. 
In most cases (72\%) the separation
($\Delta$ in pc) between neighboring knots is larger than the sum of their radii
and thus our method provides in such cases size values down to
the background level. The remaining 28\% are marked with a $\ddagger$ in 
Tables~\ref{tab5}--\ref{tab9}.

\subsection{Final sample}

\begin{figure}
\epsscale{0.88}
\plotone{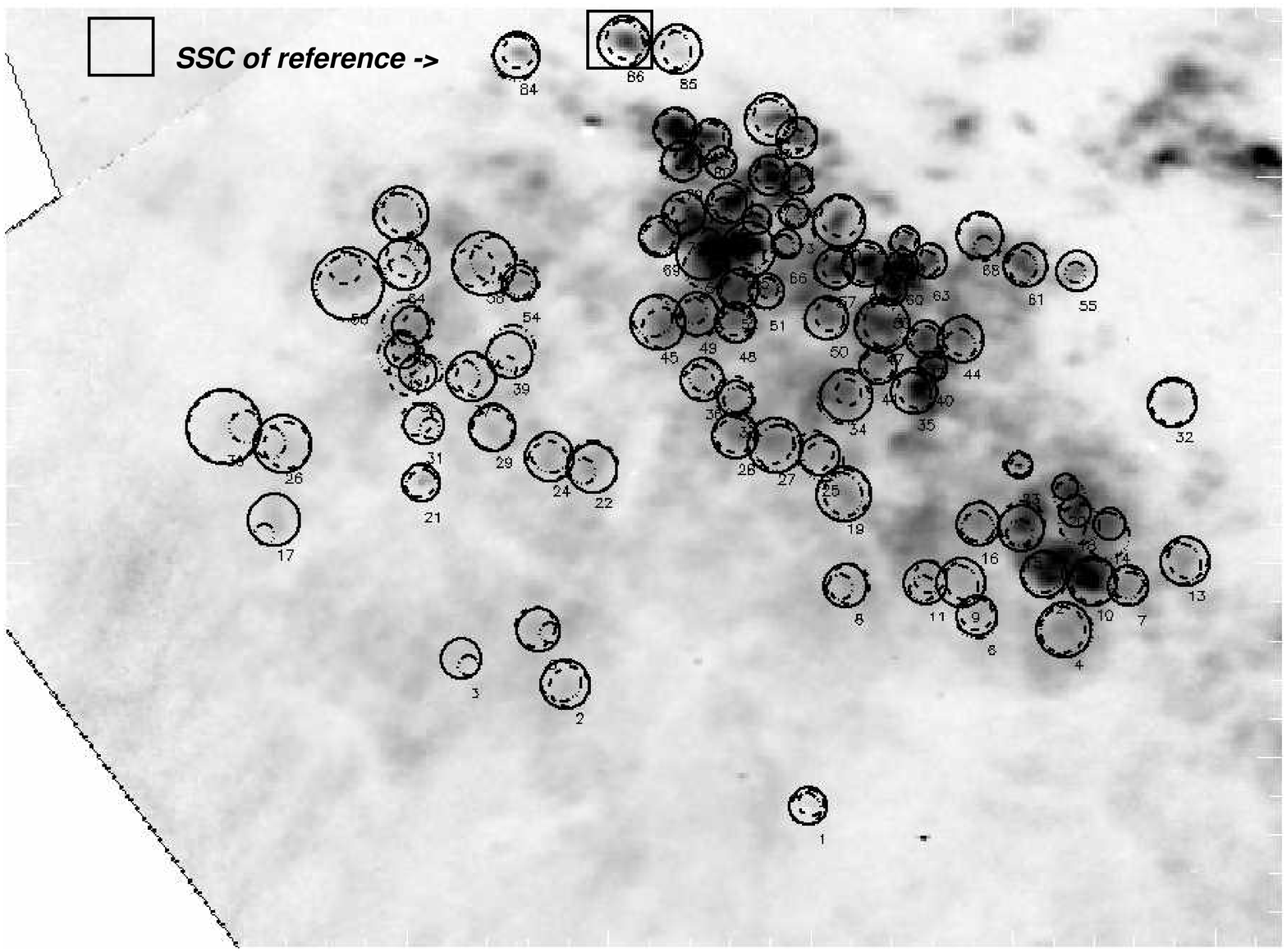}
\caption{Final sample of young SSCs within the SE zone. Young SSC
radii are overplotted on the image: the  dashed lines are H$\alpha$ knots, the
dash-dotted lines are continuum knots, and the solid lines are the final radii  
including
both types of emission. The SSC used as a reference point is indicated within a 
box. (\it
See the electronic edition of the Journal for a color version of this figure.)
\label{fig:fig3}}
\end{figure}

We compiled a catalogue of the  bright knots in H$\alpha$, 
and in the continuum images. Then, 
from both listings we  take only those that emit both in H$\alpha$ 
and the continuum and that overlap by at least half the diameter of the 
smaller determined areas. The final associated 
SSC radii (solid lines in Figure~\ref{fig:fig3}) account for the overlap
between the estimated size of the H$\alpha$ (dashed line in 
Figure~\ref{fig:fig3}) and of 
the continuum emission (dash-dotted line in Figure~\ref{fig:fig3}). 
When one of the apertures (H$\alpha$ or continuum) includes the other, we
take the larger one. In all other cases, a new aperture engulfing the other
two is defined and used as the radius of the corresponding SSCs.

\begin{figure}
\epsscale{0.35}
\plotone{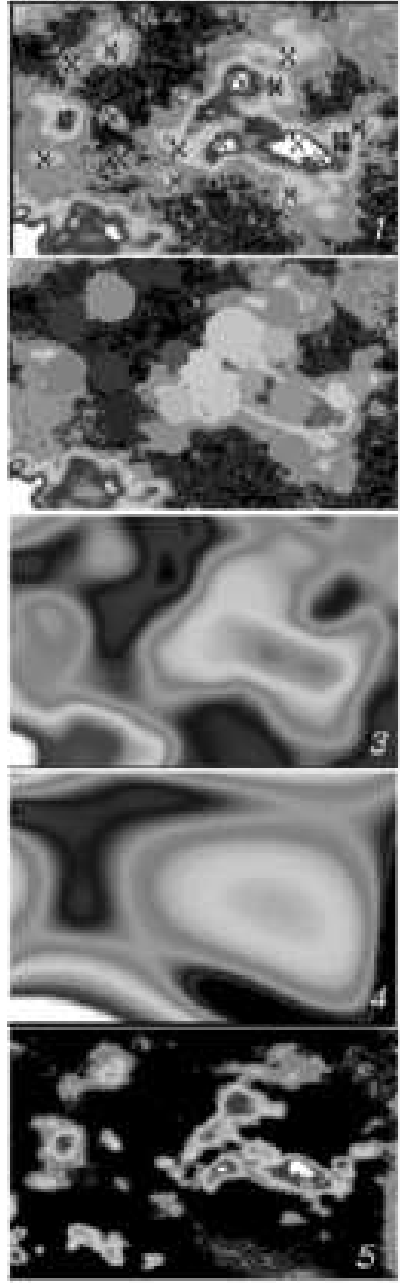}
\caption{The method. Sequence of steps followed to model and subtract the diffuse 
emission
in region M82-I. 1: H$\alpha$ image with young SSCs marked; 2: H$\alpha$ image
with overlapping masks in it (see text for further explanation); 3: map 2 is
smoothed with a Gaussian; 4: map 3 is fitted with a spline function in two
dimensions; 5: H$\alpha$ image without diffuse emission. (\it
See the electronic edition of the Journal for a color version of this figure.)
\label{fig:fig4}}
\end{figure}

We have subtracted the underlying diffuse emission of the galaxy and 
present in the catalogue the total fluxes as well as fluxes without the diffuse 
emission contribution.
Figure~\ref{fig:fig4} shows, as an example, the sequence of steps followed 
to 
model the diffuse emission map for region M82-I. The first step is to create 
masks in the H$\alpha$ map on each SSC position, of 
the same size as the radius of the SSC and with a value equal to the average 
flux found within the following annulus (1 pixel wide) after that radius 
(step~2 in Figure~\ref{fig:fig4}).
We then smooth this mask-map with a Gaussian,
the width of which is equal to the average radius for each region 
(Figure~\ref{fig:fig4}, step~3).
A model of the diffuse emission using a larger gaussian leads to border 
effects around the SSCs.
After testing several other possibilities we
finally adjusted the smoothed map with two dimensional spline functions 
\citep[FIT-REGION, IDL,][]{Molowny94}.
The best adjustments are with a 
fourth-order function (Figure~\ref{fig:fig4}, step~4).
The adjusted map is the diffuse emission map to be subtracted from the 
H$\alpha$ map in order to have a measure of the SSC
fluxes without this contribution.
Naturally, the procedure will only select as a SSC candidate
those knots with a H$\alpha$ emission larger, in each aperture, than 
its local diffuse 
emission as derived from the diffuse emission map.

We also adjusted the diffuse emission map of the continuum image 
to estimate the H$\alpha$ equivalent width ($W({\rm H}\alpha)$)
without the diffuse emission contribution. We obtained the $W({\rm H}\alpha)$ of 
the
H$\alpha$ emission line as the ratio of the H$\alpha$
emission line flux to the continuum flux, both without the diffuse emission 
contribution, within the final catalogued aperture of each young SSC.

\subsection{Comparison with FOCAS}

We used the FOCAS package in IRAF for comparison. FOCAS is basically used for 
automatic detection, photometry, and cataloguing of H~{\scshape ii} regions, 
using masks with 
the areas of the structures found. As in all methods, the problems start 
in crowded regions when the flux from neighboring bright knots begins to 
overlap.


Table~\ref{tab2} compares the fluxes measured with FOCAS and with our 
method in region M82-I. Fluxes and areas 
measured with our method are larger than those measured with FOCAS. 
The most important reason is that in most regions there are large
differences in the determined object positions between the two methods. The 
differences arise because FOCAS takes the geometrical center of regions and our 
method looks for the brightest point. FOCAS is able to find a region with 
almost any shape, while our method assumes SSCs to be compact, centrally 
concentrated round clusters and avoids contamination  with other H$\alpha$ 
emission possibly associated with them.

\subsection{Extinction, ages and masses}\label{section3.4}

	As previously mentioned, the central regions of M82 experience a
strong and variable extinction, hence each of the clusters in our sample 
requires an individualized extinction measurement. In order to calculate 
it,
we extracted the F814W, F555W, F547M and F439W fluxes\footnote{Broad-band
magnitudes are presented in the appendix (Table~\ref{tab12}). Please note that
we have included all the SSCs of all zones in the same table. The id number
(column~1) has also a reference which refers to the zone where the cluster is
measured, e.g. NE27 means source number 27, located in zone NE.} 
for each cluster 
using an aperture-photometry program written in IDL and we applied to them the
aperture correction \citep{Holtzman95}. The program was run using
the original images (unrotated and without drizzling) to minimize flux 
calibration effects. In order to measure the extinction from the continuum
colors, we used FITMODEL \citep{Maiz04}, a $\chi^2$-minimization code that 
searches through the parameter space for multiple solutions and provides 
uncertainty 
estimates with possible pre-established constraints. We inputted the 
WFPC2 photometry for all the clusters in our sample
into FITMODEL using as a comparison Starburst99 SEDs 
\citep{Leitherer99} of ages between $10^6$ and $10^{10}$ years,
with solar metallicity, and extinguished using a 
\citet{Cardelli89} law\footnote{$E(44005-5495)$ and
$R_{5495}$ are the monochromatic equivalents to $E(B-V)$ and $R_V$, 
respectively.
4405 and 5495 are the assumed central wavelengths (in \AA) of the $B$ and $V$ 
filters, respectively. Monochromatic quantities are used in FITMODEL 
because $E(B-V)$ and $R_V$ depend not only on the amount and type of 
dust but also on the model SEDs.} with $R_{5495} = 3.1$. FITMODEL generates a 
likelihood plot as a function of the chosen parameters, in our case age and
$E(4405-5495)$, that can be used to derive the expected values and their 
uncertainties.  Unfortunately, the derivation of ages and extinction of 
stellar clusters using broad-band optical photometry is hampered by 
the existence of color degeneracies for the parameters one is trying to 
measure \citep{Anders04b}. Therefore, for many clusters the reddening values 
derived in this way are of little use due to their large uncertainties unless 
some additional information about the age is included. 

	Ages can be estimated independently using the equivalent widths of
the Balmer lines. During the first 6 Myr of the life of a cluster formed in
an instantaneous burst of star formation, the existent O and WR stars 
generate large numbers of ionizing photons which produce large equivalent widths 
of the Balmer emission lines (for solar metallicity and H$\alpha$, 
$W({\rm H}\alpha)$ larger than 
100 \AA). At an age of $\approx$ 6 Myr, the last of the single O stars explodes 
as a SN and the ionizing flux of the cluster would drop to very low values if it 
was not for O stars produced in binaries by mass transfer 
\citep{Cervino99,vanBever99}. Those stars ``rejuvenate'' the cluster and 
maintain 
values of the $W({\rm H}\alpha)$ between 10 and 100 \AA\ until the cluster is
$\sim$ 25 Myr old. Unfortunately, a number 
of factors such as photon leakage, presence of an underlying stellar population,
differential extinction, and stochastic fluctuations in the IMF 
\citep[see, e.g][]{Maiz98,Cervino03} hamper the derivation of exact ages from 
the equivalent widths of the Balmer lines for unresolved or 
slightly resolved young clusters. Despite that caveat, for solar metallicities 
it is quite safe to give an age between 1 and 6 Myr for a cluster with 
$W({\rm H}\alpha) > 100$ \AA\ and an age between 6 and 25 Myr for a cluster with
$100$ \AA\ $ > W({\rm H}\alpha) > 10$ \AA. For the case of M82, the only 
circumstance
that is likely to invalidate the above criterion for age estimation would be the
existence of two clusters, one young and one old, along the same line of sight.

With the age constraints derived from the $W({\rm H}\alpha)$, we obtained the 
expected values and uncertainties for the extinction of each cluster. Results 
are shown in Tables~\ref{tab5}--\ref{tab9} (column~9). In a few cases,
the FITMODEL output indicated that the likelihood of the age ranges determined 
from the 
$W({\rm H}\alpha)$ was significantly lower than that of older ages, likely due 
to 
superposition between two clusters of different ages. Those cases should be 
considered 
to be more uncertain than the rest and are marked in 
Tables~\ref{tab5}--\ref{tab9} by a $\dagger$ symbol.

The extinction values obtained in this study are in agreement with the contour 
map extinction presented by \cite{Waller92}.

Stellar masses have been estimated from broad-band magnitudes 
(stellar continuum) taking into account the age and $W({\rm H}\alpha)$ 
constraints 
explained in the previous paragraphs and using the following expression:

\begin{equation} 2.5~\log(\frac{M}{M_\odot})~=~M_{V,1}(0) + (5~\log(d)~-~5) -
                 (m_V~-~A_V~-~C(t)) \end{equation} 

 where: \\
 - M = cluster mass \\
 - M$_{V,1}$(0) = absolute V magnitude of a cluster normalized to one solar mass
 for zero age (obtained from Starburst99)\\
 - (5~log(d)~-~5) = distance modulus\\
 - m$_V$ = apparent V magnitude\\
 - A$_V$ = extinction coefficient in V\\
 - C(t) = age correction = M$_{V,1}$(t) - M$_{V,1}$(0)\\

For comparison, masses for SSCs with $W(H\alpha)$ larger than 100 \AA\ 
have also been estimated using Starburst99 \citep{Leitherer99} and
the $W(H\alpha)$ under the assumption of coeval bursts. The agreement
is excellent; with Starburst99 and using the $W(H\alpha)$ a
mean value of 1.97~$\times$~10$^5$ M$_\odot$ with a standard deviation of 
1.80~$\times$~10$^5$ M$_\odot$ is obtained. Following the procedure explained 
above, a mean mass of 1.75~$\times$~10$^5$ M$_\odot$ with a standard deviation 
of 2.05~$\times$~10$^5$ M$_\odot$ is obtained. In 
Tables~\ref{tab5}--\ref{tab9} 
(column~14) we present the stellar masses estimated from broad-band photometry.


\section{The Catalogue} \label{section4}

In this section we present the resulting catalogue of SSCs in the nuclear region 
of M82. 
In the first set of tables we list the positions and radii of the brightest 
knots found in the H$\alpha$ (Table~\ref{tab3}) and the continuum (Table~\ref{tab4}) 
images, sorted by declination in each zone. In these tables
we present only the first ten knots, the full lists may be acquired from  
the electronic version of the paper. The typical residual pointing error in 
WFPC2 images is 0.86$''$,  $1.84''$ being the largest error seen 
\citep[][ chap.~7]{Biretta02}, it is thus 
necessary to find a point of reference in the image. 
We could not obtain the absolute astrometry for the  
clusters due to the absence of either USNO2 \citep{Monet98}
or Tycho-2 \citep{Hog00} in the field of view of these images.
We use {\it u3jv0201r}
and {\it u3jv0202r} images as reference. The coordinates of knots and young SSCs 
are relative to the coordinates of one young SSC located within the SE 
zone (number 86 in 
Table~\ref{tab5}). This one is marked in Figure~\ref{fig:fig3} and its
coordinates are: RA = 9$^{\rm h}$~55$^{\rm m}$~53.56$^{\rm s}$ and Dec.= 
69$^\circ$~40$'$~51.78$''$. 
We have checked that this young SSC emits in all the filters we have used 
and also in the near infrared bands, e.g. in \citet{Alonso03} and 
\citet*{McCrady03}
who also report this knot (MGG-1a), so we consider it a good reference point. 
We have compared our coordinates with those provided by \citet{McCrady03}
who had adjusted their astrometry to match with the positions listed in 
\citet*{Kronberg72}. The agreement with MGG-1a \citep{McCrady03} is good 
(a difference of 2'' in declination).

Tables~\ref{tab5}--\ref{tab9} list the final sample of SSCs in each 
zone sorted by declination. 
Column 1 gives the identification number of young SSCs candidates. SSCs with 
emission outside the roundish structures are marked with an asterisk. This
additional emission is either more than one SSC superposed in projection or
clusters suffering channeled outflows of ionized material.
SSCs which have problems with their age determination are marked with a $\dagger$
symbol. 
Columns 2 and 3 list the  R.A. offset (seconds)  
and the Dec.\ offset (arc seconds) with respect to the reference SSC;
column 4 gives the radius  (pc); 
column 5 gives the H$\alpha$ flux 
(10$^{-15}$~erg s$^{-1}$ cm$^{-2}$);  
column 6 gives the estimated H$\alpha$ diffuse emission flux in each SSC
aperture (10$^{-15}$~erg s$^{-1}$ cm$^{-2}$);
column 7 lists the H$\alpha$ diffuse emission flux as a percentage 
of the H$\alpha$ flux (\%);
column 8 gives the H$\alpha$ flux without diffuse emission 
(10$^{-15}$~erg s$^{-1}$ cm$^{-2}$);
column 9 lists the internal extinction (in V magnitudes);
column 10 gives the H$\alpha$ luminosity of the SSCs corrected for galactic and 
internal extinction and without the diffuse emission contribution
(10$^{38}$~erg s$^{-1}$); 
column 11 gives the projected separation to the nearest SSC ($\Delta$, pc);
column 12 lists the number of ionizing photons (10$^{49}$~s$^{-1}$)
obtained by using the expression given by \cite{Osterbrock89};
column 13 lists the H$\alpha$ equivalent width (\AA);
and 
column 14 gives the stellar mass (10$^5$~$M_\odot$).

As mentioned in section~\ref{section2}, photometric errors are very small,
less than 5$\%$. The most important source of error in the flux determination is
the aperture selection. We have estimated this error by measuring the flux in
the two adjacent apertures to the one that defines the radius (see column 4). 
The mean value of their difference is taken as the flux 
error (columns~5, 6 and 8).

Table~\ref{tab10} summarizes the densities of young SSCs, mean values and 
standard deviation for stellar masses, luminosities, radii and projected
separation to the closest SSC for each of the five zones that compose the 
nucleus of M82. In Table~\ref{tab11} star formation rate values for the 
different zones analyzed are given. The number of SSCs for each zone is also 
provided (column 2) as well as the number of them in the two considered age ranges, 
1-6 Myr and 6-25 Myr. The star formation rate has been estimated in two ways:
using the masses provided in Tables~\ref{tab5}--\ref{tab9} and from the total 
H$\alpha$ luminosity measured on each zone. SFR ($M_\odot$~yr$^{-1}$) for SSCs 
with 1-6 Myr and 6-25 Myr respectively and estimated from the mass values 
provided in Tables~\ref{tab5}--\ref{tab9} are given in columns 3 and 4. 
Columns 5, 6 and 7 give 
the total H$\alpha$ luminosity (10$^{40}$ erg s$^{-1}$) for each zone, and the 
derived values of SFR and SFR per unit area ($kpc^2$). 
Note that the values provided for the SFR on each zone, using masses estimated 
with 
FITMODEL for younger and older SSCs and H$\alpha$ luminosity (columns 3, 4 and 
6) are in good agreement.




\section{Discussion} \label{section5}

The main features of the nuclear clusters catalogued in M82 are their youth, their 
compactness (5.7~$\pm$~1.4~pc radii), their high luminosity and, therefore, 
masses and the SSCs surface density (richness). We found 197 young SSCs in 
the nuclear starburst of M82 
and more may be hidden behind the dark lanes and may have similar properties to 
those of the SSCs that we have found. \cite{Lipscy04} found seven
star-forming clusters in the mid infrared but neither of them seem to match with 
the
SSCs here found. Using radio images they proved that the mid-IR sources are
heavily obscured H~{\scshape ii} regions so these mid-IR SSCs may be hidden in the
H$\alpha$ images.

\begin{figure}
\epsscale{1}   
\plotone{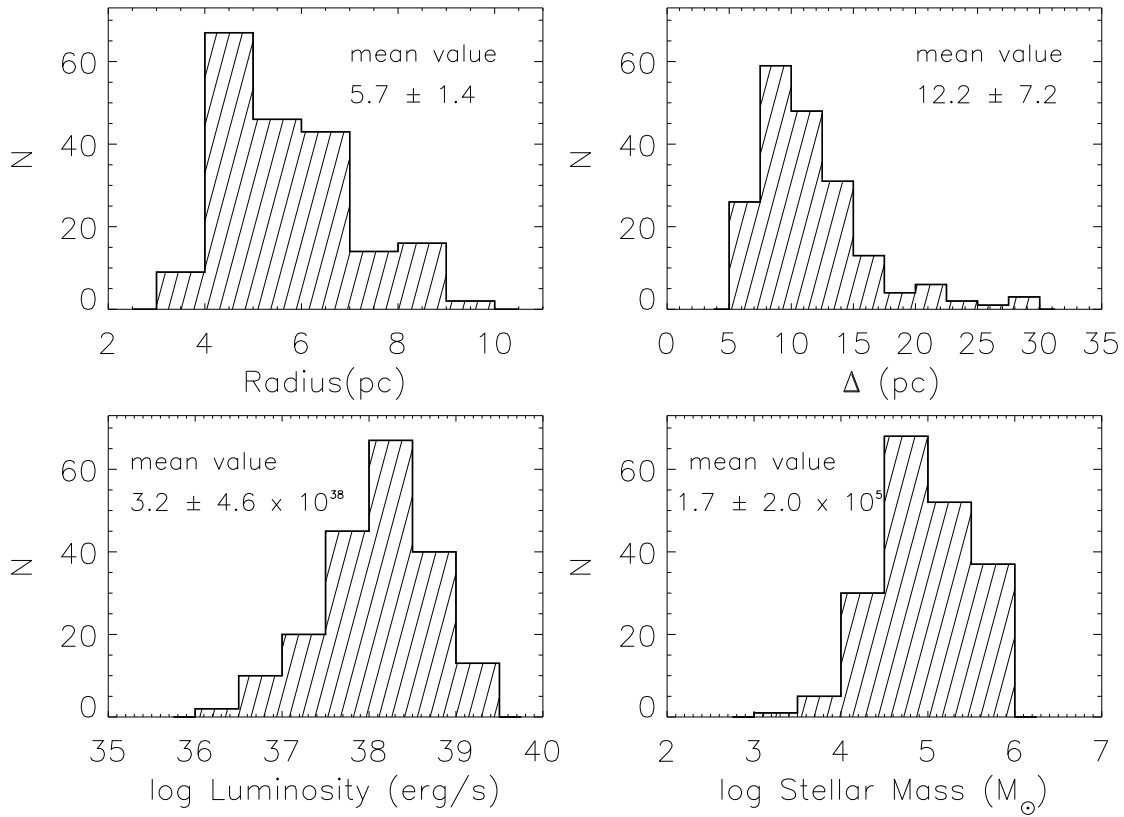}
\caption{The properties of super star clusters in the nuclear starburst of M82. 
Upper--left: histogram of the radii of SSCs (in parsecs). 
Upper--right: histogram of projected separation to the closest SSC (in parsecs). 
Bottom--left: histogram of young SSC luminosities (units: 10$^{38}$~erg s$^{-
1}$).
These values are corrected for galactic and internal extinction and without the 
diffuse emission contribution. Bottom--right: histogram of
stellar masses within young SSCs (units: 10$^{5}$~M$_\odot$).
\label{fig:fig5}}
\end{figure}

Statistics in the SSC parameters for every analyzed zone (5 in total) are
given in Table~\ref{tab10}. The analysis is summarized in the histograms shown 
in Figure~\ref{fig:fig5}.

SSCs are very compact objects; for example, \cite{Meurer95} found analysing UV
images that the typical FWHM of SSCs in a sample of galaxies is 2.6~pc. The 
WFPC2 images of M82 in H$\alpha$ had been analyzed by \cite{Oconnell95}, who 
found a typical FWHM  of 3.5~pc for SSCs. Here we have developed a method 
to estimate the physical size of SSCs that takes into account the H$\alpha$
emission, the emission in the continuum and their overlap. The latter
leads to larger size values than other methods where radii are obtained from 
lower spatial resolution images in only one filter (V, from the WF camera).
Figure~\ref{fig:fig5} (top-left) shows a histogram with the 
size distribution of SSCs in M82. Radii are in the range 3--9~pc with a mean 
value of 5.7~$\pm$~1.4~pc (Table~\ref{tab10}). 
Note however, that our values are very similar to those found by 
\cite{Billett02} for SSCs in nearby dwarf irregular galaxies and also similar
to the values inferred for SSCs in M82-F \citep{Smith01} who found a half-light
radius of 2.8 $\pm$ 0.3 pc $\sim$ 1/2 FWHM. Both of the latter studies based on
HST/WFPC2 images.

The projected separation to the closest neighbors for each of the young 
SSC is smaller than 30~pc (see Figure~\ref{fig:fig5}--top-right), with a 
minimum value 
of 5~pc and a mean separation of 12.2~$\pm$~7.2~pc (Table~\ref{tab10}). The 
projected typical separation between SSCs (see Tables~\ref{tab5}--\ref{tab9}) 
is factors of two to three times larger than the typical size associated to 
them.

The global density of young SSCs, or number of young SSCs per unit area, 
is very high
in the starburst of M82 (see Table~\ref{tab10}). \cite{Oconnell95}  
reported about 50 individual luminous clusters in M82-A; this region is 
equivalent to our SE zone, where we have catalogued 86 SSCs. Zone SE 
extends over a square area of 
242~$\times$~242~pc$^2$ and therefore has a young SSC density of 
1468~kpc$^{-2}$. For the whole starburst of M82, we estimate a SSC density of 
620~kpc$^{-2}$. If one compares with other nearby nuclear starburst, like 
NGC~253, the SSCs density of M82 is much larger. NGC~253 has only four SSCs
in its nuclear zone, within an area of 197~$\times$~180~pc \citep{Watson96} and 
thus a SCC density of 113~kpc$^{-2}$, 5.5 times smaller than M82. The 
differences become larger if the comparison is made with zones NE, N and SE 
where the highest SSCs density is found in M82.

The SSCs in the nuclear starburst of M82 also show a high H$\alpha$ luminosity
(see Figure~\ref{fig:fig5}--bottom-left). H$\alpha$ luminosities, corrected for 
galactic 
and internal extinction, and without the diffuse emission contribution are in 
the range 0.01--23~$\times$~10$^{38}$~erg~s$^{-1}$ with a mean value at
3.2~$\times$~10$^{38}$~erg~s$^{-1}$ (Table~\ref{tab10}) for all young SSCs in 
the starburst of M82. If we add together all the SSCs we obtain a total 
H$\alpha$ luminosity of 6.3~$\times$~10$^{40}$~erg~s$^{-1}$.

The stellar masses of SSCs in M82 have been determined 
using Starburst99 and FITMODEL and our results are displayed in 
Figure~\ref{fig:fig5} (bottom right-hand panel). The young SSC
mass in M82 ranges from 4~$\leq$ $log(M_{\rm SSC}/M_\odot)$ $\leq$~6 with a mean
value of 1.7~$\pm$~2.0~$\times$~10$^5$~$M_\odot$ (see 
Table~\ref{tab10}). The measured stellar masses highly depend on the applied
extinction correction. However, the most massive clusters in M82, have masses 
similar to those in M82-F, spectroscopically confirmed by \cite{Smith01}.
They are also similar to mass estimates for clusters in NGC~1569 
\citep{Anders04a} that range between 5.7~$\leq$~log(M$_{\rm SSC}/M_\odot$)~$\leq$~6.2.
On the other hand, \cite{deGrijs03b} found even more massive clusters in 
NGC~6745 reaching values up to 10$^8$ $M_\odot$, although given the distance to
this source, compact SSC may be spatially unresolved.
\cite{McCrady03} measured kinematic masses of two of the IR brightest 
clusters in M82, using high-resolution IR spectroscopy. Only one of them emits 
in H$\alpha$, MGG-11, which corresponds to our cluster number 5 in the NW zone.
Their measured kinematic mass is 3.5~$\times$~10$^5$~$M_\odot$, which,
considering the difference in the methods, compares rather well with our
determination of 4.1~$\times$~10$^5$~$M_\odot$.

The star formation rate for the different zones is in the range (0.03--0.8)
$M_\odot$ yr$^{-1}$ (Table~\ref{tab11}). The numbers are consistent after 
obtaining 
the SFR separately for younger (1-6 Myr) and older clusters (6-25 Myr). 
The agreement is also very good with the values of the SFR for each zone 
obtained using 
the total H$\alpha$ luminosity (column 5, 6 in Table~\ref{tab11}).
The star formation rate per unit area, $\Sigma_{\rm SFR}$ (see column 7, 
Table~\ref{tab11})
is very large in the five different zones. Our values, ranging between 500-4100,
are in agreement with 
values reported for starbursts by 
\cite*[][see his Figure~6]{Kennicutt98}. 
Comparing 
the values for the different zones, the lowest value corresponds to zone N. 
Although 
the SSC density is very high in this zone, the mean luminosity of the SSCs is 
the 
lowest (see Table~\ref{tab10}). 
\cite{Kennicutt98} presented also the case of 30 Doradus, which, in its inner 
10~pc, 
has $\Sigma_{\rm SFR}$~$\sim$ 100~$M_\odot$~yr$^{-1}$~kpc$^{-2}$, but the 
mean $\Sigma_{\rm SFR}$ over the entire H~{\scshape ii} region is 
$\sim$~1--10~M$_\odot$~yr$^{-1}$~kpc$^{-2}$. M82 in thus a case of exacerbated
star formation well above the average in the local Universe.

Also important is the spatial distribution of the SSCs in M82. 
As mentioned before, the largest SSC density (see Table~\ref{tab10}) is clearly 
the eastern one 
(within the zones labeled NE, N, and SE), which are right at the base of 
the supergalactic wind filamentary structure. In this respect, as pointed out by 
\cite{Tenorio03}, the interaction of the winds from neighboring SSCs is most 
probably the key ingredient in the development of a large scale filamentary 
structure embedded in a pool of X-ray emitting gas as observed in M82. Here, 
we have identified relevant observational parameters in the study of 
SGWs, such as the density of SSCs and the comparison between their typical 
sizes and projected separation. From the catalogue we find that the mean radii
of the catalogued clusters is similar to the mean distance among them, which 
seems to be a necessary condition for the individual winds to interact and 
generate the oblique shock waves able to channel the material along, 
straight structures such as the filaments seen in M82.

More detailed analysis of the cluster distribution and the base of the super 
wind structure will be the subject of a forthcoming paper. There we will use NIR
data to identify any possible SSCs completely hidden from view in the optical
data and to improve our extinction and age measurements.


\begin{acknowledgements}

We would like to thank our anonymous  referee for multiple 
comments and suggestions that led to a major revision of the
method here used to infer the SSC parameters. 
We also thank Almudena Alonso-Prieto and collaborators for allowing 
access to their data, and VM acknowledges Rosa M. 
Gonz\'alez-Delgado for her generous help with the use of Starburst99 
code. VM acknowledges support from the visitor 
program of the DDRF at STScI. The authors also acknowledge Terry Mahoney and the 
Scientific Editorial Service of the IAC for the careful checking 
of the manuscript.
This project has been partly funded by the Spanish DGC (AYA2001-3939-C03-03) 
and CONACYT-M\'exico (grant 36132-E).

\end{acknowledgements}

\clearpage

\begin{deluxetable}{lrrrr} 
\tablecolumns{4} 
\tablewidth{0pc} 
\tablecaption{Log of {\itshape HST} data archive observations} 
\tablehead{
\colhead{Filter} & \colhead{P.I.} & \colhead{Band} & \colhead{Images} & 
\colhead{Exposure times (s)} \label{tab1}}  
\startdata

F656N & W. Sparks  & H$\alpha$ & u2s04201t--2t & 2 $\times$ 300  \\
F656N & P. Shopbell& H$\alpha$ & u3jv0101r--102r--201r--202r& 4 $\times$ 500\\
F658N &            & N~{\scshape ii}        & u3jv0103r--104r--203r--204r& 4 
$\times$ 600 \\
F547M &    & Str.\ {\it y}$^a$ & u3jv0207r--8r            & 2 $\times$ 50   \\
F814W & R. O'Connell& WFPC2 I   & u45t0105r--6r--7r--8r & 2 $\times$ 
200/1200/600 \\
      &            &           & --hr--ir--jr--kr      & 260/140/2 $\times$ 900 
\\
F439W &            & WFPC2 B   & u45t0109r--ar--br--cr & 2 $\times$ 
400/1000/2300 \\
      &            &           & --lr--mr--nr--or      & 500/300/2500/1100 \\
F555W &            & WFPC2 V   & u45t0101r--2r--3r--4m & 2 $\times$ 350/1600/800 
\\
      &            &           &  --dr--er--fr--gr     & 300/200/2 $\times$ 1000 
\\

\enddata 

\tablenotetext{a}{Wide Str\"{o}mgren $y$.}

\end{deluxetable} 


\begin{deluxetable}{rrrrrrr} 
\tablecolumns{7} 
\tablewidth{0pc} 
\tablecaption{Comparison between results obtained with FOCAS
and our method for the region M82-I.} 
\tablehead{
\colhead{Id.} & \colhead{$R$$^{a}$} & \colhead{$F_{\rm H_\alpha,us}^{b}$} & 
\colhead{$F_{\rm H_\alpha,fo}^{b}$} & \colhead{$F_{\rm us}/F_{\rm fo}$} & 
\colhead{$A_{\rm us}/A_{\rm fo}^{c}$} & \colhead{$\delta$$^d$} \label{tab2}}  
\startdata
1  &  5.64 & 4.88  &  3.76 & 1.30 & 1.59& 3.2 \\
2  &  5.64 & 2.84  &  1.28 & 2.22 & 1.68& 1.8 \\
3  &  5.64 & 7.13  &  4.94 & 1.44 & 1.43& 1.8 \\
4  &  4.03 & 16.33 &  2.35 & 6.96 & 8.75& 2.6 \\
5  &  5.64 & 9.59  &  1.31 & 7.34 & 7.33& 1.7 \\
6  &  6.44 & 29.47 & 22.85 & 1.29 & 1.59& 2.0 \\
7  &  7.25 & 54.24 & 22 96 & 2.47 & 3.44& 2.5 \\
8  &  4.03 & 10.00 &  6.54 & 1.53 & 1.68& 2.3 \\
9  &  4.03 & 5.89  &  2.96 & 1.99 & 2.32& 1.5 \\
10 &  4.03 & 11.53 &  1.38 & 8.35 & 9.85& 2.7 \\
11 &  6.44 & 17.34 & 12.52 & 1.38 & 1.62& 2.1 \\
12 &  6.44 & 4.10  &  2.37 & 1.73 & 1.85& 0.8 \\
13 &  6.44 & 23.07 &  3.66 & 6.31 &10.09& 0.9 \\
14 &  5.64 & 10.95 &  1.04 &10.58 &15.38& 4.3 \\
15 &  8.86 & 45.42 & 20.43 & 2.22 & 3.56& 2.6 \\
16 &  6.44 & 4.62  &  1.71 & 2.70 & 3.30& 1.4 \\
17 &  5.64 & 7.66  &  3.66 & 2.09 & 2.14& 1.0 \\
18 &  7.25 & 11.57 &  3.26 & 3.55 & 6.07& 4.4 \\

\enddata 
\tablenotetext{a}{Radii of SSCs obtained with our method, in parsecs.}
\tablenotetext{b}{Fluxes in 10$^{-15}$~erg s$^{-1}$ cm$^{-2}$.}
\tablenotetext{c}{Area ratio of both methods (our method/FOCAS).}
\tablenotetext{d}{Differences in the determined object positions between the two
methods, in pixels.}

\end{deluxetable} 


\clearpage
\pagestyle{empty}

\begin{deluxetable}{lrrrrrrrrrrrrrrrrrrr} 
\tabletypesize{\footnotesize}
\rotate
\tablecolumns{20} 
\tablewidth{0pc} 
\tablecaption{\footnotesize H$_\alpha$ bright knots for each zone. 
Coordinates of knots are relative to the reference SSC 
(RA = 9$^{\rm h}$~55$^{\rm m}$~53.56$^{\rm s}$ and 
Dec. = 69$^\circ$~40$'$~51.78$''$) and they are sorted by declination.
[$\Delta$RA units: seconds; $\Delta$Dec. units: arc seconds; Radius units:
parsecs.]} 
\tablehead{
\colhead{} & \multicolumn{3}{c}{N zone} & \colhead{} &  \multicolumn{3}{c}{NE 
zone} &
\colhead{} & \multicolumn{3}{c}{NW zone} & \colhead{} &  \multicolumn{3}{c}{SE 
zone} &
\colhead{} & \multicolumn{3}{c}{SW zone} \\
\cline{2-4} \cline{6-8} \cline{10-12} \cline{14-16} \cline{18-20} \\
\colhead{Id.}    &   \colhead{$\Delta$RA}   &  \colhead{$\Delta$Dec.} & 
\colhead{$R$}  &
\colhead{}    &   \colhead{$\Delta$RA}   &  \colhead{$\Delta$Dec.} & 
\colhead{$R$}  &
\colhead{}    &   \colhead{$\Delta$RA}   &  \colhead{$\Delta$Dec.} &
\colhead{$R$}  &
\colhead{}    &   \colhead{$\Delta$RA}   &  \colhead{$\Delta$Dec.} & 
\colhead{$R$}  &
\colhead{}    &   \colhead{$\Delta$RA}   &  \colhead{$\Delta$Dec.} & 
\colhead{$R$}  
\label{tab3}}  
\startdata

  1& $-$2.261 &$-$0.70& 3.22&& $-$1.821 & $-$2.85 & 4.03&& $-$3.684 & $-$7.14 & 
5.64 & &  $-$0.072& $-$13.02 & 4.03 && $-$1.513 & $-$18.34 & 7.25  \\
  2& $-$2.266 &$-$0.30& 3.22&& $-$1.704 & $-$2.54 & 4.03&& $-$3.089 & $-$6.96 & 
5.64 & &    0.097 & $-$12.44 & 4.83 && $-$1.697 & $-$18.33 & 4.83  \\
  3& $-$2.171 &  0.10 & 4.83&& $-$1.814 & $-$2.40 & 2.42&& $-$3.610 & $-$6.81 & 
4.83 & &  $-$0.030& $-$11.99 & 6.44 && $-$1.520 & $-$17.92 & 4.83  \\
  4& $-$2.582 &  0.16 & 4.03&& $-$1.305 & $-$2.34 & 4.83&& $-$2.709 & $-$6.70 & 
4.03 & &  $-$0.381& $-$11.70 & 5.64 && $-$1.563 & $-$17.37 & 5.64  \\
  5& $-$2.323 &  0.39 & 4.03&& $-$1.117 & $-$1.93 & 4.83&& $-$3.292 & $-$6.40 & 
4.83 & &    0.144 & $-$11.68 & 6.44 && $-$1.655 & $-$17.36 & 5.64  \\
  6& $-$2.451 &  0.62 & 4.03&& $-$1.815 & $-$1.90 & 4.83&& $-$3.210 & $-$6.40 & 
4.83 & &    0.855 & $-$11.39 & 4.83 && $-$1.458 & $-$17.10 & 2.42  \\
  7& $-$2.259 &  0.66 & 4.03&& $-$0.849 & $-$1.90 & 4.83&& $-$2.685 & $-$6.08 & 
6.44 & &    1.060 & $-$11.34 & 5.64 && $-$1.652 & $-$16.67 & 6.44  \\
  8& $-$2.388 &  1.02 & 4.03&& $-$1.313 & $-$1.73 & 3.22&& $-$2.806 & $-$5.84 & 
4.03 & &  $-$0.343& $-$11.29 & 3.22 && $-$2.332 & $-$16.62 & 4.03  \\
\enddata 
\\

\tablecomments{The complete version of this table is in the electronic edition 
of
the Journal. The printed edition contains only a sample.}

\end{deluxetable} 


\begin{deluxetable}{lrrrrrrrrrrrrrrrrrrr} 
\tabletypesize{\footnotesize}
\rotate
\tablecolumns{20} 
\tablewidth{0pc} 
\tablecaption{\footnotesize Continuum bright knots for each zone. 
Coordinates of knots are relative to the reference SSC 
(RA = 9$^{\rm h}$~55$^{\rm m}$~53.56$^{\rm s}$ and 
Dec. = 69$^\circ$~40$'$~51.78$''$) and they are sorted by declination.
[$\Delta$RA units: seconds; $\Delta$Dec. units: arc seconds; Radius units
parsecs.]} 
\tablehead{
\colhead{} & \multicolumn{3}{c}{N zone} & \colhead{} &  \multicolumn{3}{c}{NE 
zone} &
\colhead{} & \multicolumn{3}{c}{NW zone} & \colhead{} &  \multicolumn{3}{c}{SE 
zone} &
\colhead{} & \multicolumn{3}{c}{SW zone} \\
\cline{2-4} \cline{6-8} \cline{10-12} \cline{14-16} \cline{18-20} \\
\colhead{Id.}    &   \colhead{$\Delta$RA}   &  \colhead{$\Delta$Dec.} & 
\colhead{$R$}  &
\colhead{}    &   \colhead{$\Delta$RA}   &  \colhead{$\Delta$Dec.} & 
\colhead{$R$}  &
\colhead{}    &   \colhead{$\Delta$RA}   &  \colhead{$\Delta$Dec.} & 
\colhead{$R$}  &
\colhead{}    &   \colhead{$\Delta$RA}   &  \colhead{$\Delta$Dec.} & 
\colhead{$R$}  &
\colhead{}    &   \colhead{$\Delta$RA}   &  \colhead{$\Delta$Dec.} & 
\colhead{$R$}  
\label{tab4}}  
\startdata

  1& $-$2.235 &$-$0.65& 3.22&& $-$1.720 & $-$2.69 & 4.83&& $-$3.674 & $-$7.08 & 
6.44 & &$-$0.486& $-$11.23& 4.03 && $-$3.201 & $-$16.45 & 5.64    \\
  2& $-$2.257 &$-$0.35& 3.22&& $-$1.278 & $-$2.37 & 4.83&& $-$3.506 & $-$6.79 & 
3.22 & &  0.171 & $-$9.37 & 4.83 && $-$3.730 & $-$16.16 & 7.25    \\
  3& $-$2.375 &$-$0.15& 4.83&& $-$1.111 & $-$2.06 & 4.83&& $-$3.319 & $-$6.78 & 
4.83 & &$-$1.319& $-$9.26 & 2.42 && $-$3.987 & $-$15.59 & 4.83    \\
  4& $-$2.161 &  0.08 & 3.22&& $-$1.801 & $-$2.00 & 3.22&& $-$2.663 & $-$6.00 & 
4.83 & &  0.423 & $-$9.14 & 2.42 && $-$2.397 & $-$15.31 & 3.22    \\
  5& $-$2.559 &  0.13 & 3.22&& $-$0.847 & $-$1.95 & 4.83&& $-$2.469 & $-$5.43 & 
5.64 & &$-$0.687& $-$8.85 & 6.44 && $-$2.282 & $-$14.94 & 5.64    \\
  6& $-$2.293 &  0.31 & 3.22&& $-$1.297 & $-$1.85 & 4.83&& $-$2.742 & $-$4.89 & 
3.22 & &  0.151 & $-$8.74 & 4.83 && $-$3.245 & $-$14.67 & 5.64    \\
  7& $-$2.256 &  0.57 & 2.42&& $-$1.675 & $-$1.77 & 7.25&& $-$3.093 & $-$4.65 & 
3.22 & &  0.102 & $-$8.71 & 3.22 && $-$2.194 & $-$14.30 & 6.44    \\
  8& $-$2.330 &  0.70 & 3.22&& $-$1.454 & $-$1.68 & 5.64&& $-$3.171 & $-$4.62 & 
4.83 & &  0.204 & $-$8.66 & 2.42 && $-$3.197 & $-$13.94 & 5.64    \\
\enddata 
\\
\tablecomments{The complete version of this table is in the electronic edition 
of
the Journal. The printed edition contains only a sample.}

\end{deluxetable} 


\clearpage

\pagestyle{plain}


\begin{deluxetable}{lrrrccccccrcrc} 
\tabletypesize{\footnotesize}
\rotate
\tablecolumns{14} 
\tablewidth{22cm}
\setlength{\tabcolsep}{0.05in} 
\tablecaption{Candidate young SSCs in the N zone.} 
\tablehead{
\colhead{Id.}    &   \colhead{$\Delta$RA}   &  \colhead{$\Delta$Dec.}   & 
\colhead{$R$} & \colhead{F$_{\rm H\alpha}$} 
& \colhead{$F_{\rm diff}$} & \colhead{$\%$$F_{\rm H\alpha}$} & \colhead{$F_{\rm 
H\alpha}'$} 
& \colhead{$A_V$} & \colhead{L$_{\rm H\alpha}$} &
\colhead{$\Delta$}   & \colhead{$N_{\rm fot}^a$}&
 \colhead{W($H_\alpha$)} & \colhead{$M_\star$}   \\
\colhead{(1)}&\colhead{(2)}& 
\colhead{(3)}&\colhead{(4)}&\colhead{(5)}&\colhead{(6)}&
\colhead{(7)}&\colhead{(8)}& 
\colhead{(9)}&\colhead{(10)}&\colhead{(11)}&\colhead{(12)}&
\colhead{(13)}&\colhead{(14)}
 \label{tab5}  }
\startdata

  1$^*$$^\ddagger$& $-$2.250&$-$0.68&3.22&  2.6$\pm$1.1 & 1.50$\pm$0.75& 58 & 
1.11$\pm$0.32 & 4.67$\pm$0.40& 0.63 $\pm$0.37 & 6.24&  4.6 $\pm$ 2.7 & 102 
&0.241$\pm$0.080 \\
  2 	          & $-$2.168 & 0.09 &4.03&  3.7$\pm$1.2 &  3.2$\pm$1.3 & 87 
&0.459$\pm$0.084& 4.15$\pm$0.70& 0.18 $\pm$0.13 &11.37&  1.29$\pm$ 0.92&  54 
&0.55 $\pm$0.50  \\
  3$^*$           & $-$2.572 & 0.14 &4.83&  6.1$\pm$1.5 & 2.70$\pm$0.94& 44 & 
3.42$\pm$0.59 & 4.68$\pm$0.35& 1.95 $\pm$0.85 &24.17& 14.3 $\pm$ 6.2 & 123 &0.46 
$\pm$0.17  \\
  4$^*$$^\ddagger$& $-$2.310 & 0.35 &4.03& 10.4$\pm$3.5 &  7.1$\pm$2.8 & 69 & 
3.25$\pm$0.74 & 3.66$\pm$0.23& 0.86 $\pm$0.34 & 6.62&  6.3 $\pm$ 2.5 & 122 &0.40 
$\pm$0.11  \\
  5$^\ddagger$    & $-$2.259 & 0.61 &3.22&  4.9$\pm$2.3 &  4.0$\pm$2.0 & 81 & 
0.92$\pm$0.29 & 3.59$\pm$0.49& 0.23 $\pm$0.16 & 6.62&  1.7 $\pm$ 1.2 &  42 &0.65 
$\pm$0.57  \\
  6$^*$$^\ddagger$& $-$2.267 & 1.05 &3.22&  4.4$\pm$2.0 &  3.7$\pm$1.8 & 84 & 
0.69$\pm$0.15 & 4.22$\pm$0.18& 0.28 $\pm$0.10 & 7.61&  2.03$\pm$ 0.73& 109 &0.47 
$\pm$0.11  \\
  7$^*$           & $-$2.715 & 1.73 &8.86& 20.2$\pm$2.8 & 11.1$\pm$1.9 & 55 & 
9.12$\pm$0.91 & 3.92$\pm$0.47& 2.9  $\pm$1.3  &15.73& 21.5 $\pm$ 9.7 &  82 &1.9  
$\pm$1.5   \\
  8 	          & $-$2.683 & 2.78 &8.05& 19.1$\pm$3.7 & 14.7$\pm$2.9 & 77 & 
4.39$\pm$0.78 & 4.37$\pm$0.75& 2.0  $\pm$1.5  &13.22& 15.  $\pm$11.  & 192 &0.68 
$\pm$0.48  \\
\enddata 
\tablecomments{See section~\ref{section5} for explanations of 
Tables~\ref{tab5}~-~\ref{tab9}.
Coordinates of knots are relative to the reference SSC 
(RA = 9$^{\rm h}$~55$^{\rm m}$~53.56$^{\rm s}$ and 
Dec. = 69$^\circ$~40$'$~51.78$''$) and they are sorted by declination.
[$\Delta$RA units: seconds; $\Delta$Dec. units: arc seconds; 
Flux units: 10$^{-15}$~erg s$^{-1}$ cm$^{-2}$; 
Luminosity units: 10$^{38}$~erg s$^{-1}$;  
N$_{fot}$ units: 10$^{49}$~s$^{-1}$; Mass units: 10$^{5}$~$M_\odot$].
$^a$~$Q(H^o)~=~7.31~\cdot~10^{11}~L_{\rm H_\alpha}$ (\cite{Osterbrock89}). }
\end{deluxetable} 



\begin{deluxetable}{lrrrccccccrcrc} 
\tabletypesize{\footnotesize}
\rotate
\tablecolumns{15} 
\tablewidth{22cm} 
\setlength{\tabcolsep}{0.05in} 
\tablecaption{Candidate young SSCs in the NE zone.\label{tab6}} 
\tablehead{
\colhead{Id.}    &   \colhead{$\Delta$RA}   &  \colhead{$\Delta$Dec.}   & 
\colhead{$R$} & \colhead{F$_{\rm H\alpha}$} 
& \colhead{$F_{\rm diff}$} & \colhead{$\%$$F_{\rm H\alpha}$} & \colhead{$F_{\rm 
H\alpha}'$} 
& \colhead{$A_V$} & \colhead{L$_{\rm H\alpha}$} &
\colhead{$\Delta$}   & \colhead{$N_{\rm fot}^a$}&
 \colhead{W($H_\alpha$)} & \colhead{$M_\star$}  \\
\colhead{(1)}&\colhead{(2)}& 
\colhead{(3)}&\colhead{(4)}&\colhead{(5)}&\colhead{(6)}&
\colhead{(7)}&\colhead{(8)}& 
\colhead{(9)}&\colhead{(10)}&\colhead{(11)}&\colhead{(12)}&
\colhead{(13)}&\colhead{(14)}
  }
\startdata

  1$^*$$^\ddagger$    	& $-$1.293 & $-$2.35 &5.64 & 4.88$\pm$0.95 &  
3.4$\pm$1.0  & 70 &1.469$\pm$0.063 &6.18$\pm$0.47 &  2.6  $\pm$ 1.1  & 9.88& 
19.2 $\pm$ 7.7 &  36 &	$\S$	  \\
  2        	  	& $-$1.116 & $-$1.99 &5.64 & 2.84$\pm$0.54 & 2.62$\pm$0.79 & 
93 &  0.21$\pm$0.25 &6.06$\pm$0.90 &  0.35 $\pm$ 0.64 &14.28&  2.5 $\pm$ 4.7 &  
25 &	$\S$	   \\
  3$^*$$^\ddagger$    	& $-$1.810 & $-$1.95 &4.03 & 16.3$\pm$5.1  &  
6.2$\pm$2.4  & 38 & 10.2$\pm$2.6   &5.09$\pm$0.17 &  8.0  $\pm$ 3.1  & 8.19& 58.  
$\pm$22.  & 116 &2.19 $\pm$0.47  \\
  4        	  	& $-$0.850 & $-$1.93 &5.64 &  7.1$\pm$1.2  & 3.10$\pm$0.95 &
44 &  4.03$\pm$0.29 &4.68$\pm$0.18 &  2.32 $\pm$ 0.48 &13.35& 16.9 $\pm$ 3.5 & 
123 &0.435$\pm$0.089\\
  5$^*$    	  	& $-$1.307 & $-$1.79 &5.64 &  9.6$\pm$2.5  &  6.7$\pm$1.9  & 
70 &  2.90$\pm$0.64 &6.18$\pm$0.47 &  5.2  $\pm$ 3.0  & 9.88& 38.  $\pm$22.  &  
32 &9.5  $\pm$7.2   \\
  6$^*$$^\ddagger$    	& $-$1.706 & $-$1.71 &7.25 & 54.2$\pm$7.5  & 
21.6$\pm$4.7  & 40 & 32.6$\pm$2.8   &4.59$\pm$0.14 & 17.6  $\pm$ 3.4  
&10.37&128.  $\pm$25.  & 130 &5.7  $\pm$1.0  \\
  7$^*$    	  	& $-$1.462 & $-$1.66 &6.44 & 29.5$\pm$4.6  & 14.2$\pm$3.5  &
48 & 15.3$\pm$1.1   &5.92$\pm$0.18 & 22.3  $\pm$ 4.6  &11.22&163.  $\pm$34.  & 
130 &6.6  $\pm$1.3   \\
  8$^\ddagger$        	& $-$1.873 & $-$1.62 &4.03 & 10.0$\pm$2.6  &  
4.8$\pm$1.9  & 48 &  5.16$\pm$0.68 &4.96$\pm$0.21 &  3.7  $\pm$ 1.1  & 6.81& 
26.7 $\pm$ 7.8 & 106 &1.25 $\pm$0.28  \\
\enddata 
\tablecomments{See section~\ref{section5} for explanations of 
Tables~\ref{tab5}~-~\ref{tab9}.
Coordinates of knots are relative to the reference SSC 
(RA = 9$^{\rm h}$~55$^{\rm m}$~53.56$^{\rm s}$ and 
Dec. = 69$^\circ$~40$'$~51.78$''$) and they are sorted by declination.
[$\Delta$RA units: seconds; $\Delta$Dec. units: arc seconds; 
Flux units: 10$^{-15}$~erg s$^{-1}$ cm$^{-2}$; 
Luminosity units: 10$^{38}$~erg s$^{-1}$;  
N$_{fot}$ units: 10$^{49}$~s$^{-1}$; Mass units: 10$^{5}$~$M_\odot$].
$^a$~$Q(H^o)~=~7.31~\cdot~10^{11}~L_{\rm H_\alpha}$ (\cite{Osterbrock89}). 
$\S$ We do not give mass values for these SSCs because we do not have 
broad-band magnitudes for them.}

\end{deluxetable} 



\begin{deluxetable}{lrrrccccccrcrc} 
\tabletypesize{\footnotesize}
\rotate
\tablecolumns{15} 
\tablewidth{22cm} 
\setlength{\tabcolsep}{0.05in} 
\tablecaption{Candidate young SSCs in the NW zone.} 
\tablehead{
\colhead{Id.}    &   \colhead{$\Delta$RA}   &  \colhead{$\Delta$Dec.}   & 
\colhead{$R$} & \colhead{F$_{\rm H\alpha}$} 
& \colhead{$F_{\rm diff}$} & \colhead{$\%$$F_{\rm H\alpha}$} & \colhead{$F_{\rm 
H\alpha}'$} 
& \colhead{$A_V$} & \colhead{L$_{\rm H\alpha}$} &
\colhead{$\Delta$}   & \colhead{$N_{\rm fot}^a$}&
 \colhead{W($H_\alpha$)} & \colhead{$M_\star$}  \\
\colhead{(1)}&\colhead{(2)}& 
\colhead{(3)}&\colhead{(4)}&\colhead{(5)}&\colhead{(6)}&
\colhead{(7)}&\colhead{(8)}& 
\colhead{(9)}&\colhead{(10)}&\colhead{(11)}&\colhead{(12)}&
\colhead{(13)}&\colhead{(14)}
 \label{tab7}  }
\startdata

  1$^*$       	 	    & $-$3.681 & $-$7.11& 6.44 & 14.3$\pm$2.3 &  
8.8$\pm$2.2 & 61 &5.561$\pm$0.091 &4.48$\pm$0.73 & 2.7 $\pm$ 1.6 & 62.96 & 20. 
$\pm$11. & 90&1.09$\pm$0.87  \\
  2 	      	 	    & $-$2.675 & $-$6.04& 6.44 & 11.0$\pm$2.2 &  
6.1$\pm$1.5 & 55 &  4.98$\pm$0.67 &5.82$\pm$0.83 & 6.8 $\pm$ 5.2 & 21.19 & 50. 
$\pm$38. & 64&4.0 $\pm$3.3   \\
  3$^*$       	 	    & $-$3.101 & $-$4.63& 4.83 & 13.9$\pm$3.3 &  
6.4$\pm$2.1 & 46 &  7.6$\pm$1.2   &5.16$\pm$0.47 & 6.3 $\pm$ 3.2 & 14.43 & 46. 
$\pm$23. & 66&4.1 $\pm$3.1   \\
  4 	      	 	    & $-$2.708 & $-$4.20& 7.25 & 25.9$\pm$3.3 & 
14.6$\pm$3.2 & 56 &11.301$\pm$0.048&5.22$\pm$0.47 & 9.7 $\pm$ 3.5 & 12.64 & 71.
$\pm$25. & 90&5.9 $\pm$4.7   \\
  5$^*$       	 	    & $-$3.324 & $-$4.03& 5.64 & 14.2$\pm$3.1 &  
9.7$\pm$2.8 & 68 &  4.55$\pm$0.38 &5.67$\pm$0.49 & 5.5 $\pm$ 2.5 & 14.86 & 40. 
$\pm$18. & 58&4.1 $\pm$3.3   \\
  6$^*$$^\ddagger$          & $-$3.043 & $-$3.86& 7.25 & 58.5$\pm$7.0 & 
18.6$\pm$4.1 & 32 & 39.9$\pm$2.9   &3.99$\pm$0.44 &13.6 $\pm$ 5.5 & 11.81 & 99.
$\pm$40. & 81&8.6 $\pm$6.3   \\
  7$^*$$^\dagger$$^\ddagger$& $-$2.622 & $-$3.52& 4.83 & 23.8$\pm$5.3 &  
6.3$\pm$2.1 & 27 & 17.5$\pm$3.3   &5.80$\pm$0.16 &23.4 $\pm$ 7.2 &  8.69 &171. 
$\pm$53. &125&4.14$\pm$0.87   \\
  8$^*$$^\ddagger$          & $-$3.604 & $-$3.49& 4.83 &  9.9$\pm$2.6 &  
6.4$\pm$2.1 & 64 &  3.55$\pm$0.48 &5.87$\pm$0.21 & 5.0 $\pm$ 1.5 &  7.51 & 37. 
$\pm$11. &177&1.64$\pm$0.35   \\
\enddata 
\tablecomments{See section~\ref{section5} for explanations of 
Tables~\ref{tab5}~-~\ref{tab9}.
Coordinates of knots are relative to the reference SSC 
(RA = 9$^{\rm h}$~55$^{\rm m}$~53.56$^{\rm s}$ and 
Dec. = 69$^\circ$~40$'$~51.78$''$) and they are sorted by declination.
[$\Delta$RA units: seconds; $\Delta$Dec. units: arc seconds; 
Flux units: 10$^{-15}$~erg s$^{-1}$ cm$^{-2}$; 
Luminosity units: 10$^{38}$~erg s$^{-1}$;  
N$_{fot}$ units: 10$^{49}$~s$^{-1}$; Mass units: 10$^{5}$~$M_\odot$].
$^a$~$Q(H^o)~=~7.31~\cdot~10^{11}~L_{\rm H_\alpha}$ (\cite{Osterbrock89}). }

\end{deluxetable} 


\begin{deluxetable}{lrrrccccccrcrc} 
\centering
\tabletypesize{\footnotesize}
\rotate
\tablecolumns{15} 
\tablewidth{22cm} 
\setlength{\tabcolsep}{0.05in} 
\tablecaption{Candidate young SSCs in the SE zone.\label{tab8}} 
\tablehead{
\colhead{Id.}    &   \colhead{$\Delta$RA}   &  \colhead{$\Delta$Dec.}   & 
\colhead{$R$} & \colhead{F$_{\rm H\alpha}$} 
& \colhead{$F_{\rm diff}$} & \colhead{$\%$$F_{\rm H\alpha}$} & \colhead{$F_{\rm 
H\alpha}'$} 
& \colhead{$A_V$} & \colhead{$L_{\rm H\alpha}$} &
\colhead{$\Delta$}   & \colhead{$N_{\rm fot}^a$}&
 \colhead{W($H_\alpha$)} &  \colhead{$M_\star$}   \\
\colhead{(1)}&\colhead{(2)}& 
\colhead{(3)}&\colhead{(4)}&\colhead{(5)}&\colhead{(6)}&
\colhead{(7)}&\colhead{(8)}& 
\colhead{(9)}&\colhead{(10)}&\colhead{(11)}&\colhead{(12)}&
\colhead{(13)}&\colhead{(14)}
 }  
\startdata

  1 	 	 	    &$-$0.498&$-$11.19& 4.83 &  4.5$\pm$1.4 &  3.7$\pm$1.2 & 
82&  0.83$\pm$0.21&1.35$\pm$0.47 &0.038$\pm$0.023& 44.36& 0.28$\pm$ 0.17&  13 & 
0.091$\pm$0.067	\\
  2 	 	 	    &  0.157& $-$9.42 & 6.44 & 14.6$\pm$3.5 &  8.7$\pm$2.2 & 
60&  5.9$\pm$1.3  &2.34$\pm$0.31 &0.58 $\pm$0.26 & 15.41& 4.2 $\pm$ 1.9 & 106 & 
0.097$\pm$0.035	\\
  3 	 	 	    &  0.440& $-$9.04 & 4.83 &  7.8$\pm$2.5 &  5.2$\pm$1.7 & 
66&  2.67$\pm$0.78&1.65$\pm$0.49 &0.16 $\pm$0.10 & 20.53& 1.14$\pm$ 0.75&  66 & 
0.090$\pm$0.081	\\
  4$^*$  	 	    & $-$1.189&$-$8.63& 6.44 & 28.2$\pm$6.6 & 19.1$\pm$4.8 & 
68&  9.0$\pm$1.8  &2.64$\pm$0.53 &1.11 $\pm$0.67 & 14.52& 8.1 $\pm$ 4.9 &  99 & 
0.37 $\pm$0.20	\\
  5 	 	 	    &  0.230& $-$8.62 & 5.64 & 14.9$\pm$3.9 &  6.2$\pm$1.8 & 
42&  8.7$\pm$2.1  &2.58$\pm$0.28 &1.02 $\pm$0.46 & 15.41& 7.5 $\pm$ 3.4 & 111 & 
0.116$\pm$0.030	\\
  6$^*$  	 	    & $-$0.953&$-$8.41& 4.83 & 11.4$\pm$3.4 & 10.9$\pm$3.6 & 
95&  0.54$\pm$0.20&3.53$\pm$0.46 &0.130$\pm$0.092&  9.24& 0.95$\pm$ 0.68&  20 & 
1.4  $\pm$1.2    \\
  7$^*$$^\ddagger$  	    & $-$1.359&$-$7.98& 4.83 & 17.2$\pm$5.2 & 
11.3$\pm$3.7 & 66&  5.9$\pm$1.5  &3.77$\pm$0.49 &1.7  $\pm$1.1  &  8.63&12.4 
$\pm$ 7.7 &  86 & 1.08 $\pm$0.95    \\
  8$^*$$^\ddagger$  	    & $-$0.596&$-$7.96& 5.64 & 16.6$\pm$4.4 & 
13.0$\pm$3.7 & 78&  3.59$\pm$0.72&3.05$\pm$0.25 &0.60 $\pm$0.23 &  8.51& 4.4 
$\pm$ 1.7 & 126 & 0.273$\pm$0.081  \\
\enddata 

\tablecomments{See section~\ref{section5} for explanations of 
Tables~\ref{tab5}~-~\ref{tab9}.
Coordinates of knots are relative to the reference SSC 
(RA = 9$^{\rm h}$~55$^{\rm m}$~53.56$^{\rm s}$ and 
Dec. = 69$^\circ$~40$'$~51.78$''$) and they are sorted by declination.
[$\Delta$RA units: seconds; $\Delta$Dec. units: arc seconds; 
Flux units: 10$^{-15}$~erg s$^{-1}$ cm$^{-2}$; 
Luminosity units: 10$^{38}$~erg s$^{-1}$;  
N$_{fot}$ units: 10$^{49}$~s$^{-1}$; Mass units: 10$^{5}$~$M_\odot$].
$^a$~$Q(H^o)~=~7.31~\cdot~10^{11}~L_{\rm H_\alpha}$ (\cite{Osterbrock89}). 
$\S$ We do not give mass values for these SSCs because we do not have 
broad-band magnitudes for them.}
\end{deluxetable} 



\begin{deluxetable}{lrrrccccccrcrc} 
\tabletypesize{\footnotesize}
\rotate
\tablecolumns{15} 
\tablewidth{22cm}
\setlength{\tabcolsep}{0.05in} 
\tablecaption{Candidate young SSCs in the SW zone.} 
\tablehead{
\colhead{Id.}    &   \colhead{$\Delta$RA}   &  \colhead{$\Delta$Dec.}   & 
\colhead{$R$} & \colhead{F$_{\rm H\alpha}$} 
& \colhead{$F_{\rm diff}$} & \colhead{$\%$$F_{\rm H\alpha}$} & \colhead{$F_{\rm 
H\alpha}'$} 
& \colhead{$A_V$} & \colhead{L$_{\rm H\alpha}$} &
\colhead{$\Delta$}   & \colhead{$N_{\rm fot}^a$}&
 \colhead{W($H_\alpha$)} & \colhead{$M_\star$}  \\
 \colhead{(1)}&\colhead{(2)}& 
\colhead{(3)}&\colhead{(4)}&\colhead{(5)}&\colhead{(6)}&
\colhead{(7)}&\colhead{(8)}& 
\colhead{(9)}&\colhead{(10)}&\colhead{(11)}&\colhead{(12)}&
\colhead{(13)}&\colhead{(14)}
\label{tab9}  }
\startdata

  1 	         	    & $-$3.758& $-$16.14& 8.05 &18.8$\pm$3.6 & 13.0$\pm$2.6 & 
69 &  5.8$\pm$1.0  & 3.14$\pm$0.52 &1.04 $\pm$0.59 & 59.45& 7.6  $\pm$ 4.3  & 
142 &0.231$\pm$0.097     \\
  2$^*$          	    & $-$2.384& $-$15.30& 5.64 &12.6$\pm$3.3 & 10.3$\pm$2.9 & 
82 &  2.31$\pm$0.38& 2.11$\pm$0.45 &0.190$\pm$0.097& 11.93& 1.39 $\pm$ 0.71 &  
48 &0.20 $\pm$0.14  \\
  3 	         	    & $-$2.274& $-$14.94& 6.44 &14.9$\pm$3.6 & 14.5$\pm$3.6 & 
97 &0.383$\pm$0.065& 1.64$\pm$0.54 &0.022$\pm$0.013& 11.93& 0.162$\pm$ 0.094&   
7 &0.119$\pm$0.090 \\
  4 	         	    & $-$2.172& $-$14.31& 7.25 &23.1$\pm$4.5 & 20.1$\pm$4.5 & 
87 &2.991$\pm$0.025& 2.30$\pm$0.67 &0.28 $\pm$0.15 & 14.30& 2.1  $\pm$ 1.1  &  
51 &0.28 $\pm$0.22  \\
  5$^*$          	    & $-$3.216& $-$11.82& 8.86 & 95.$\pm$15. &  59.$\pm$11. & 
62 & 35.8$\pm$4.5  & 2.51$\pm$0.63 &4.0  $\pm$2.4  & 15.75&29.   $\pm$18.   & 
253 &0.38 $\pm$0.25      \\
  6$^*$$^\dagger$	    & $-$2.968& $-$11.76& 8.05 & 75.$\pm$13. &  53.$\pm$11. & 
71 & 21.8$\pm$2.5  & 2.41$\pm$0.17 &2.25 $\pm$0.54 & 18.54&16.   $\pm$ 3.9  & 
440 &0.41 $\pm$0.10      \\
  7 	         	    & $-$2.616& $-$11.22& 6.44 & 49.$\pm$13. &  35.$\pm$8.7 & 
70 & 14.6$\pm$3.9  & 3.06$\pm$0.21 &2.5  $\pm$1.0  & 16.89&18.0  $\pm$ 7.6  & 
175 &0.56 $\pm$0.16      \\
  8$^*$          	    & $-$3.122& $-$11.07& 8.05 & 94.$\pm$17. &  54.$\pm$11. & 
58 & 39.2$\pm$6.1  & 2.93$\pm$0.76 &6.0  $\pm$4.4  & 15.75&44.   $\pm$32.   & 
477 &0.47 $\pm$0.35      \\
\enddata 
\tablecomments{See section~\ref{section5} for explanations of 
Tables~\ref{tab5}~-~\ref{tab9}.
Coordinates of knots are relative to the reference SSC 
(RA = 9$^{\rm h}$~55$^{\rm m}$~53.56$^{\rm s}$ and 
Dec. = 69$^\circ$~40$'$~51.78$''$) and they are sorted by declination.
[$\Delta$RA units: seconds; $\Delta$Dec. units: arc seconds; 
Flux units: 10$^{-15}$~erg s$^{-1}$ cm$^{-2}$; 
Luminosity units: 10$^{38}$~erg s$^{-1}$;  
N$_{fot}$ units: 10$^{49}$~s$^{-1}$; Mass units: 10$^{5}$~$M_\odot$].
$^a$~$Q(H^o)~=~7.31~\cdot~10^{11}~L_{\rm H_\alpha}$ (\cite{Osterbrock89}). }

\end{deluxetable} 


\begin{deluxetable}{rcrcccccccccccc} 
\tabletypesize{\footnotesize}
\tablecolumns{15} 
\rotate
\setlength{\tabcolsep}{0.04in} 
\tablewidth{0pc} 
\tablecaption{\footnotesize Density (number of SSCs per unit area, kpc$^{-2}$) 
of young SSCs 
within each zone. Mean (~$\bar{}$~) and median values (~$\widehat{}$~) and
their standard deviation ($\sigma$) for stellar masses (10$^5$~$M_\odot$), 
H$_\alpha$ luminosities (10$^{38}$~erg~s$^{-1}$), radius (pc) and projected 
separation to the closest SSC (pc) for each zone. } 
\tablehead{
\colhead{Zone}& \colhead{Size (pc)}  &\colhead{$\rho$}   & 
\colhead{$\overline{M}$} &
\colhead{$\widehat{M}$} & \colhead{$\sigma$({\scshape M})} & 
\colhead{$\overline{L}$} & 
\colhead{$\widehat{L}$} & \colhead{$\sigma$({\scshape L})} & 
\colhead{$\overline{R}$} & 
\colhead{$\widehat{R}$} & \colhead{$\sigma$({\scshape R})} & 
\colhead{$\overline{\Delta}$} &
\colhead{$\widehat{\Delta}$}&\colhead{$\sigma({\Delta})$}  
  \label{tab10}}
\startdata
N    &161.0 $\times$ 64.4 &1640&0.53&0.43&0.50 &0.98&0.69&0.89&5.6&5.6&1.8&11.2&10.9&4.4\\
NE   &201.3 $\times$ 120.8&2097&1.8 &1.0 &2.1  & 3.4& 1.4& 5.2&5.8&5.6&1.3&11.5&10.8&4.2 \\
NW   &185.2 $\times$ 120.8& 939&4.0 &4.1 &2.0  & 9.5& 6.8& 7.6&6.3&5.6&1.5&15  &12  &12  \\
SE   &242.0 $\times$ 242.0&1468&1.7 &0.9 &2.0  & 2.0& 1.5& 2.1&5.3&4.8&1.2&10.7&9.4 &5.1 \\
SW   &281.8 $\times$ 241.5& 338&0.65&0.38&0.88 & 2.6& 2.2& 2.3&6.4&6.4&1.3&18  &15  &12 \\
Total&563.5 $\times$ 563.5& 620&1.7 &0.9 &2.0  & 3.2& 1.7& 4.6&5.7&5.6&1.4&12.2&10.5&7.2 \\
  
\enddata 

\end{deluxetable} 


\begin{deluxetable}{rcccccc} 
\tabletypesize{\footnotesize}
\tablecolumns{7} 
\rotate
\setlength{\tabcolsep}{0.04in} 
\tablewidth{0pc} 
\tablecaption{\footnotesize Star formation rate values for the different 
zones analyzed. Columns 2, 3 and 4 provide the number the SSCs (in 
parentesis the number of them with 1-6 Myr and with 6-25 Myr), SFR 
($M_\odot$~yr$^{-1}$) for SSCs with 1-6 Myr and 6-25 Myr respectively and 
estimated from the mass values provided in Tables~\ref{tab5}--\ref{tab9}.
Columns 5, 6 and 7 
give the total H$\alpha$ luminosity (10$^{40}$ erg s$^{-1}$) for each zone,
and the derived values of SFR and SFR per unit area ($kpc^2$).
} 
\tablehead{
\colhead{Zone}& \colhead{$N$} &\colhead{$SFR_j$}  &\colhead{$SFR_o$}   & 
\colhead{$L_{H\alpha}$} &\colhead{$SFR_{H\alpha}$} & \colhead{$\Sigma_{SFR}$}  
\\
\colhead{} &\colhead{(Ny+No)} &\colhead{\scriptsize{1-6 Myr}} &
\colhead{\scriptsize{6-25 Myr}} &\colhead{}
&\colhead{}   &\colhead{}   
\label{tab11}}
\startdata
N    & 16(8+8)    &0.055& 0.031&0.45& 0.033& 508    \\
NE   & 51(16+35)  &0.77 & 0.27 & 3.6& 0.26 & 4014   \\
NW   & 21(6+15)   &0.52 & 0.30 & 3.8& 0.28 & 3285   \\
SE   & 86(26+60)  &0.31 & 0.66 & 4.8& 0.35 & 4112   \\
SW   & 23(17+6)   &0.16 & 0.034& 1.7& 0.13 & 1502   \\
Total& 197(73+124)&1.8  & 1.3  & 14 & 1.0  & 2731   \\
  
\enddata 

\end{deluxetable} 


\clearpage

\acknowledgments

\appendix 
 \section{APPENDIX: BROAD-BAND MAGNITUDES}

\begin{deluxetable}{lcccc} 
\tabletypesize{\scriptsize}
\tablecolumns{5} 
\tablewidth{0pc} 
\tablecaption{\footnotesize \label{tab12}} 
\tablehead{
\colhead{Id.} &   \colhead{m$_{F439W}$}&\colhead{m$_{F555W}$} & 
\colhead{m$_{F814W}$}&\colhead{m$_{F547M}$}
}  
\startdata

 N01 &22.58 $\pm$ 0.28& 21.34 $\pm$ 0.26& 19.19 $\pm$ 0.12& 21.2 $\pm$ 2.0   \\
 N02 &22.67 $\pm$ 0.26& 20.98 $\pm$ 0.23& 19.08 $\pm$ 0.33& 21.0 $\pm$ 2.4   \\
 N03 &22.19 $\pm$ 0.15& 20.64 $\pm$ 0.13& 18.73 $\pm$ 0.19& 20.8 $\pm$ 2.6   \\
 N04 &20.991$\pm$0.065& 19.79 $\pm$ 0.11& 18.32 $\pm$ 0.14& 19.48$\pm$0.46   \\
 N05 &21.448$\pm$0.045& 20.01 $\pm$ 0.11& 18.43 $\pm$ 0.17& 20.07$\pm$0.57   \\
 N06 &21.602$\pm$0.044& 20.153$\pm$0.086& 18.520$\pm$0.095& 20.43$\pm$0.83   \\
 N07 &20.899$\pm$0.158& 19.30 $\pm$ 0.12& 17.413$\pm$0.073& 20.0 $\pm$ 3.5   \\
 N08 &21.51 $\pm$ 0.40& 19.91 $\pm$ 0.34& 18.25 $\pm$ 0.39& 20.3 $\pm$ 4.4   \\
\enddata
\\

\end{deluxetable} 


\clearpage


\clearpage

\clearpage

\clearpage

\clearpage

\clearpage

\end{document}